\documentclass[%
aip,
rsi, 
amsmath,amssymb,
reprint,%
]{revtex4-1}

\usepackage{graphicx}
\usepackage{dcolumn}
\usepackage{bm}

\usepackage{amsmath}
\usepackage{ amssymb }

\usepackage{float} 
\usepackage{wrapfig} 

\usepackage[makeroom]{cancel} 

\newcommand{\beq}{\begin{equation}}
\newcommand{\eeq}{\end{equation}}
\newcommand{\pa}[2]{\frac{\partial #1}{\partial #2}}
\newcommand{\vv}{\mathbf{v}}

\newcommand{\bvec}{\begin{pmatrix}}
\newcommand{\evec}{\end{pmatrix}}
\newcommand{\lp}{\left(}
\newcommand{\rp}{\right)}

\newcommand{\Bvec}{\mathbf{B}}

\newcommand{\jvec}{\mathbf{j}}

\newcommand{\tdt}{\tilde{t}}
\newcommand{\tR}{\tilde{R}}
\newcommand{\tr}{\tilde{r}}

\newcommand{\tv}{\tilde{v}}

\newcommand{\ta}{\tilde{a}}

\newcommand{\bvo}{\bar{v}_0}
\newcommand{\sgn}{\text{sgn}}

\usepackage[normalem]{ulem} 

\usepackage{comment}

\usepackage{hyperref}
\hypersetup{
    colorlinks=true,
    linkcolor=blue,
    filecolor=magenta,      
    citecolor=cyan,
    pdftitle={Current channel evolution in ideal Z pinch for general velocity profiles},
    bookmarks=true,
    hyperindex = true
}

\usepackage[titletoc]{appendix}
\usepackage{outlines}

\begin{document}

\title[Current channel evolution in ideal Z pinch for general velocity profiles]{Current channel evolution in ideal Z pinch for general velocity profiles}

\author{I. E. Ochs}
\affiliation{Department of Astrophysical Sciences, Princeton University, Princeton, NJ 08540.}
\author{C. Stollberg}
\affiliation{Weizmann Institute of Science, Rehovot 7610001, Israel}
\author{E. Kroupp}
\affiliation{Weizmann Institute of Science, Rehovot 7610001, Israel}
\author{Y. Maron}%
\affiliation{Weizmann Institute of Science, Rehovot 7610001, Israel}
\author{A. Fruchtman}%
\affiliation{Department of Physics, Holon Institute of Technology, Holon 58102, Israel}
\author{E. J. Kolmes}
\affiliation{Department of Astrophysical Sciences, Princeton University, Princeton, NJ 08540.}
\author{M. E. Mlodik}
\affiliation{Department of Astrophysical Sciences, Princeton University, Princeton, NJ 08540.}
\author{N. J. Fisch}%
\affiliation{Department of Astrophysical Sciences, Princeton University, Princeton, NJ 08540.}

\date{\today}



\begin{abstract}
	Recent diagnostic advances in gas-puff Z pinches at the Weizmann Institute for the first time allow the reconstruction of the current flow as a function of time and radius. 
	These experiments show an unexpected radially-outward motion of the current channel,  as the plasma moves radially-inward [C. Stollberg, Ph.D thesis, Weizmann Institute, 2019]. 
 	In this paper, a mechanism that could explain this current evolution is described.
	We examine the impact of advection on the distribution of current in a cylindrically symmetric plasma. 
	In the case of metric compression, $|v_r| \propto r$, the current enclosed between each plasma fluid element and the axis is conserved, and so the current profile maintains its shape.
	We show that for more general velocity profiles, this simple behavior quickly breaks down, allowing for non-conservation of current in a compressing conductor, rapid redistribution of the current density, and even for the formation of reverse currents.
	In particular, a specific inward radial velocity profile is shown to result in radially-outward motion of the current channel, recovering the surprising current evolution discovered at the Weizmann Institute.
\end{abstract}

\pacs{Valid PACS appear here}
\keywords{Suggested keywords}
\maketitle

\section{Introduction}

For many years, Z pinch experiments had limited diagnostic capability.
These limitations were particularly pronounced for the magnetic field diagnostics, where it was only possible to track the total current flowing through the system.
Recently, the development of spectroscopic diagnostics based on Zeeman splitting have enabled the reconstruction of temporally- and radially resolved magnetic  field profiles, thus yielding the evolution of the current density distribution during implosion\cite{davara1998spectroscopic} and stagnation\cite{rosenzweig2017measurements}. 

Davara et al. \cite{davara1998spectroscopic} confirmed that during the implosion phase of a fast Z pinch the entire current flows through the compressing plasma and obeys the normal diffusion assuming Spitzer resistivity. 
A very different result was found during stagnation by Rosenzweig et al.\cite{rosenzweig2017measurements}. 
Here, the proportion of current flowing through the stagnating plasma was at most a few percent.
This finding agrees with the results reported in Ref. \cite{maron2013balance}, where  it was shown that in significantly disparate experiments the magnetic  field effect on the pressure and energy balance at stagnation is negligible, leading to the conclusion that at most  1/3 of the load current flows in the stagnating plasma.

A recent spectroscopic investigation\cite{stollberg2019phd} on a different Z pinch experiment, also performed with unprecedented spatial and temporal
resolution throughout implosion and stagnation, verified the previous results which held over most of the axial length of the pinch.
However, in this study, a remarkable phenomenon was found in the column portion near the cathode.
While at the beginning of the stagnation, most of the current flowed within the small radius of the stagnating plasma, the current quickly escaped to much larger-radius as the stagnating plasma continued to compress.
This effect was not seen in the rest of the plasma column; at those positions the current was never observed to penetrate to small radii, consistent with the conclusions in Refs.~\cite{rosenzweig2017measurements, maron2013balance}.
In order to interpret these experimental results it is critical to understand the  behavior of the current distribution during a plasma implosion.

In a cylindrically symmetric system, where for all quantities $\partial / \partial \theta = \partial / \partial z = 0$, the field evolution consists of two parts: resistive diffusion and magnetic induction.
Historically, the theoretical focus has been on the former.
For instance, one of the more successful theoretical predictions of the current distribution in a Z pinch is the inverse skin effect\cite{haines1959inverse, culverwell1989nature}, a purely resistive effect which assumes no plasma motion.
In essence, the calculation showed that the only resistive solution consistent with a decreasing total current in the conductor was one with an inverse current running at the boundary.
Although such a model is sufficient and highly successful for plasmas where the resistive diffusion dwarfs the induction, as experiments grow hotter and faster, the induction effects will eventually dominate.

In cases where the induction effects have been examined, they are often in the context of specific shock solutions, and the change in field across the shock\cite{lee2000reversed}.
However, in hot systems and systems with finite resistivity, the current channel is likely to be less localized, and thus examining the effects of induction for more extended velocity profiles is necessary.

There is an intuitive notion on how a current channel compresses, which is based on the case of metric compression, where $v_r \propto r$.
In this case, the divergence of the velocity is constant within the cylinder.
Everywhere, conservation of mass implies that the density increases according to $n(t) = n(0) (R(0)/R(t))^2$, and conservation of magnetic flux implies that the magnetic field increases according to $B(t) = B(0) (R(0)/R(t))$.
Because $I \propto r B$, the magnetic field evolution implies that the enclosed current at any given point is a conserved quantity as well, $dI/dt = 0$.
Thus, the current channel smoothly compresses, maintaining its shape and conserving the total current.

In this paper, we show that for more general velocity profiles, this intuition quickly breaks down.
Flux conservation does not in general imply conservation of the total current, and the induction equation thus allows for rapid redistribution of the current density, and for the formation of reverse currents.

In Section~\ref{sec:analytic}, we rewrite the induction equation in terms of the enclosed current, showing how it takes a special form in the case of metric compression that leads to current conservation.
By considering power-law velocity profiles, we show how metric compression provides a natural boundary between regions of increasing and decreasing enclosed current.
We solve the induction equation analytically for the case of a compressing, bounded conductor, showing how a non-uniformly contracting conductor does not have a conserved global current.
In Section~\ref{sec:numerical}, we exploit the conserved quantity, the magnetic flux, to easily find numerical solutions for more complex velocity profiles, and use our intuition from the analytic solutions to understand the redistribution of the current channel in several experimentally relevant scenarios.
In Section~\ref{sec:experiment}, we compare our results to the Weizmann experiment, demonstrating qualitative agreement for the current channel expansion and re-contraction.
In Section~\ref{sec:haines}, we further show that the observed behavior of the current channel cannot be explained by the historically successful model of Haines, which considers only the resistive evolution of the field, neglecting the plasma motion.
Finally, in Section~\ref{sec:observables}, we discuss useful observables with which future experiments can more quantitatively distinguish resistive and advective effects.


\section{Current Profile Evolution in Ideal MHD} \label{sec:analytic}

We model the pinch as a cylindrically-symmetric ideal (superconducting) plasma.
In ideal MHD, the magnetic field evolves according to the induction equation:
\beq
	\pa{\Bvec}{t} = \nabla \times (\vv_e \times \Bvec). \label{eq:induction}
\eeq
The relevant velocity here is the electron velocity $\vv_e$, whereas the relevant dynamical velocity $\vv$ in MHD is mass-weighted. 
Throughout this paper, we will assume that the various species (electrons, ions, and neutrals) are collisionally equilibrated, which requires the timescales of momentum equilibration to be much shorter than the dynamical timescales of the implosions considered.
Then $\vv_e \approx \vv$ to high precision, and the frozen-in law holds even in the presence of ionization events.

With this assumption, in cylindrical coordinates, Eq.~(\ref{eq:induction}) becomes:
\begin{align}
	\pa{B_\theta}{t} &= [\nabla \times (v_r B_\theta \hat{z})]_\theta\\
	&= -\pa{}{r} \lp  v_r B_\theta \rp. \label{eq:cylInd}
\end{align}

We can relate this magnetic field to the current contained within a cylinder of radius $r$, $I(r) = 2\pi \int_{0}^{r} r j_z(r) dr$, by Ampere's Law:
\beq
	\mu_0 I(r) = 2\pi r B_\theta (r).
\eeq
By taking a partial derivative with respect to time and inserting Eq. (\ref{eq:cylInd}), we find:
\beq
	\frac{dI}{dt} \equiv \pa{I}{t} + v_r \pa{I}{r} = - I r \pa{}{r} \lp \frac{v_r}{r} \rp. \label{eq:dIdt}
\eeq

Although this is just a recasting of the induction equation in terms of the enclosed current, it says something fairly unintuitive: although magnetic field \emph{lines} are advected along with the plasma, the current is, in general, \emph{not}.
The reason that this is unintuitive is that intuitions about how the current evolves are often formed by considering ``self-similar'' or ``metric'' compression, where the divergence of the velocity is uniform across the plasma.
This metric compression is given by
\beq
	v_{r,\text{met}} \equiv v_0 \lp \frac{r}{a_0} \rp; \label{eq:metricCompress}
\eeq
indeed, this is the form chosen for the velocity by Haines \cite{haines1959inverse} when he extends his analysis to a compressing conductor.
But when $v_r$ is given by Eq.~(\ref{eq:metricCompress}), the term on the RHS of Eq.~(\ref{eq:dIdt}) disappears.

This disappearance has two implications.
First, during metric compression, the current density \emph{is} advected with the velocity.
Second, during non-metric compression, there are additional effects on the current evolution which cannot be understood within the normal intuitive framework of metric compression, and which depend on the degree of deviation from metric compression.

\begin{figure*}
	\includegraphics[width=\linewidth]{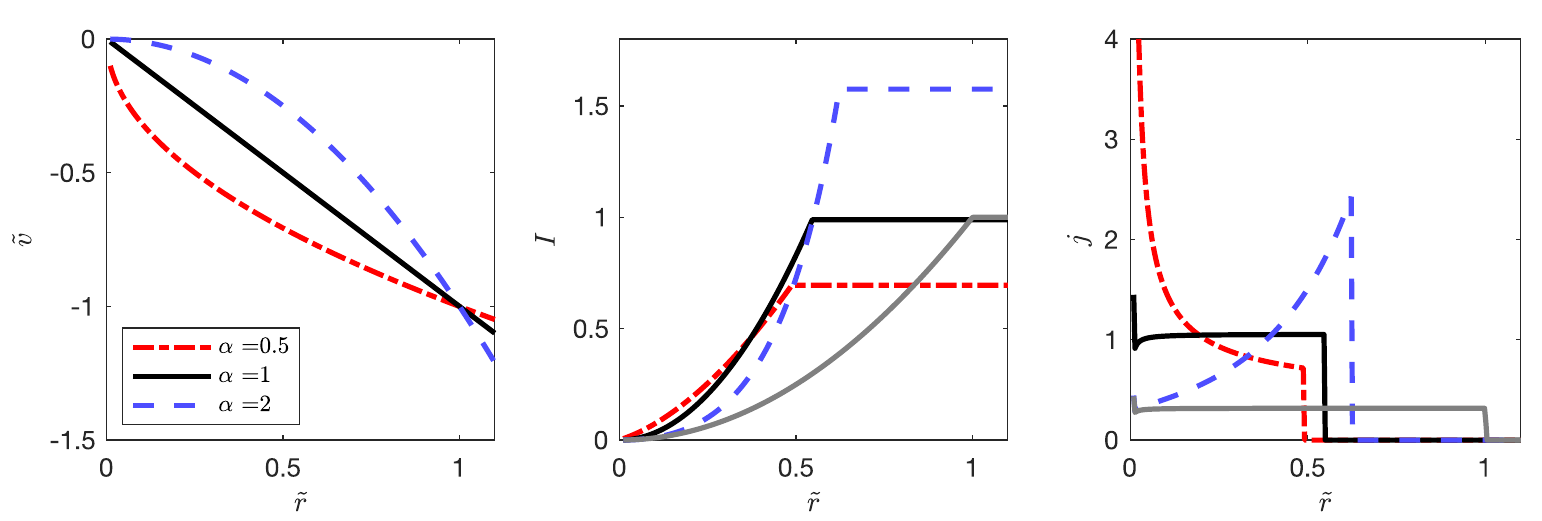}
	\caption{
		Compression of a current channel in a bounded conductor in the lab frame.
		Left: Velocity profiles sub-metric ($\alpha < 1$, red, dot-dashed), metric ($\alpha = 1$, black, solid), and super-metric ($\alpha > 1$, blue, dashed) compression.
		Center: the solution to $I(\tr,\tdt=0.6)$ for a bounded conductor with initial radius $\ta = 1$, and an initial uniform current profile, for each of the velocity profiles.
		The initial enclosed current distribution is shown in gray.
		The boundary of the conductor corresponds to the kink in the enclosed current profile.
		Right: the corresponding current density $j(\tr,\tdt=0.6)$.}
	\label{fig:boundedConductor}
\end{figure*}

\subsection{Developing intuition through analytic solutions}

To get a sense for the effects of non-metric compression, it is useful to examine the behavior for specific velocity profiles.
In particular, consider a power-law family of profiles:
\beq
	v_r (r) = v_0 \lp \frac{r}{a_0} \rp^\alpha. \label{eq:powerLawV}
\eeq
We see that $\alpha = 1$ reduces to Eq.~(\ref{eq:metricCompress}), and thus describes metric compression (or expansion).
We will describe velocity profiles with $\alpha > 1$ as ``super-metric compression'' (or expansion), and with $\alpha < 1$ as ``sub-metric compression''.

A nice feature of these power-law velocity profiles is that they allow us to analytically solve the induction equation.
To do this, we first must solve for the motion of a fluid element, which mathematically provides the characteristic curve along which Eq.~(\ref{eq:dIdt}) is solved.

It will make things cleaner to normalize this equation to a characteristic radius $\tr \equiv r/a_0$ and time $\tdt \equiv t/\tau$, yielding a characteristic nondimensional velocity $\tv = v_r / (a_0/\tau)$. 
Here, $a_0$ is a typical spatial scale of the experiment, e.g.~the radius of the outer boundary of the plasma, and $\tau$ is a typical time scale for the compression.
Then Eq.~(\ref{eq:powerLawV}) becomes:
\beq
	\tv_r (\tr) = \bvo \tr^\alpha, \label{eq:powerLawVtilde}
\eeq
where $\bvo = v_0 \tau / a_0$.
Then, for a given fluid element, the normalized radius $\tR \equiv R/a_0$ as a function of time is governed by:
\begin{align}
\frac{d\tR}{d\tdt} = \tv_r(\tR) = \bvo \tR^\alpha.
\end{align}
The solution to this equation is
\beq
	\tR = \begin{cases}
		\tR_0 e^{\bvo \tdt} & \alpha = 1\\
		\lp \tR_0^{1-\alpha} + (1-\alpha) \bvo \tdt \rp^{1/(1-\alpha)} & \alpha \neq 1,
	\end{cases}
\eeq
where $\tR_0 = \tR(\tdt=0)$.

Now, we wish to plug this fluid element motion into the induction equation, Eq.~(\ref{eq:dIdt}).
First, we plug Eq.~(\ref{eq:powerLawV}) into Eq.~(\ref{eq:dIdt}), and then we nondimensionalize, yielding
\begin{align}
	\frac{dI}{d\tdt} &= - I \tr \pa{}{\tr} \lp \bvo \tr^{\alpha-1} \rp|_{\tr=\tR}\\
	&= -(\alpha-1) \bvo \tR^{\alpha - 1} I.
\end{align}

Without yet having solved this equation, we can nevertheless gain insight into its behavior.
Specifically, the current enclosed by a moving fluid element increases or decreases according to
\beq
	\sgn \lp \frac{d|I|}{dt} \rp = - \sgn(v_r) \sgn(\alpha-1).
\eeq
Thus, the current enclosed by a fluid element in a compressing plasma ($v_r < 0$) undergoing submetric compression ($\alpha < 1$) will decrease over time.
Switching from compression to expansion or from sub- to super-metric compression will reverse this conclusion.
The full set of possibilities are laid out in Table \ref{tab:dIdt}.

\begin{table}[b]
	\begin{center}
		\begin{tabular}{ | l | c  c | } 
			\hline
			& $\bvo<0$ & $\bvo>0$\\
			\hline
			$\alpha > 1$ & $d|I|/dt > 0$ & $d|I|/dt < 0$ \\
			$\alpha < 1$ & $d|I|/dt < 0$ & $d|I|/dt > 0$ \\
			\hline
		\end{tabular}
		\caption{Evolution of the current $|I|$ enclosed by a fluid element for super-metric ($\alpha > 1$) and sub-metric ($\alpha < 1$) velocity profiles, for both compression ($\bvo<0$) and expansion ($\bvo>0$).}
		\label{tab:dIdt}
	\end{center}
\end{table}

The full solution for $I(\tr,\tdt)$ is given by
\begin{align}
	I(\tr,\tdt) &= I_0(\tR_0) \lp \frac{\tr}{\tR_0} \rp^{1-\alpha} \label{eq:Ialpha1}\\
	\tR_0(\tr,\tdt) &= \begin{cases} 
	\tr e^{-\bvo \tdt} & \alpha = 1\\
	\lp \tr^{1-\alpha} - (1-\alpha) \bvo \tdt \rp^{1/(1-\alpha)} & \alpha \neq 1. \label{eq:Ialpha4}
	\end{cases}
\end{align}
Note here that we consider $\tR_0$ to be a function of $\tr$ and $\tdt$, with the interpretation that $\tR_0(\tr_1,\tdt_1)$ represents the initial position (at $\tdt=0$) of the fluid element that is at position $\tR(\tdt_1)=\tr_1$ at time $\tdt_1$.
This definition links the Lagrangian frame of the fluid element to the fixed Eulerian coordinates of the lab frame.

These power-law solutions, when considered globally, are not particularly physical.
For $\alpha < 1$, we find finite enclosed current at $r=0$, indicating the formation of a current singularity at $\tr = 0$.
Meanwhile, for $\alpha > 1$, a singularity propagates inwards from $\tr = \infty$.
Nevertheless, they are useful for informing our view of how the magnetic field should evolve in local regions undergoing different types of compression.

\subsection{Conductor with a hard boundary}

We are now in a position to examine the most common model for the $Z$ pinch: the bounded conductor.
This model forms the conceptual basis for the snowplow and slug-piston models, as well as Haines' study of the resistive evolution of current densities\cite{haines1959inverse}.

We will study the specific case of a conductor that, at time $\tdt = 0$, extends from $\tr =0$ to $\tr = 1$, and is surrounded by a vacuum.
For this scenario, Eqs.~(\ref{eq:Ialpha1}-\ref{eq:Ialpha4}) apply for all $\tr < \ta$, where the normalized outer radius $\ta$ of the conductor is defined by $\tR_0(\ta,\tdt) = 1$.
Outside of this radius, $I$ maintains the same value as at the surface of the conductor, since the vacuum region can contain no current.
This solution is shown in Fig.~\ref{fig:boundedConductor}.

We can see instantly that the \emph{total current} within the conductor is not conserved for $\alpha \neq 1$.
Indeed, Eq.~(\ref{eq:Ialpha1}) can be used to derive an expression for the total current as a function of time, yielding
\beq
	\frac{I_{tot}(\tdt)}{I_{tot}(0)} = \ta^{1-\alpha}.
\eeq
Thus we see that, for a bounded conductor, the relationships in Table~\ref{tab:dIdt} apply not just to the local enclosed current, but to the total enclosed current as well.
Interestingly, this implies that the specific radial profile of the compression, i.e. the shape of the snowplow or slug, can have a massive impact on how much total current ends up flowing through the plasma.
Therefore, the velocity profile likely can have a large influence on the inductance of the plasma and thus the circuit dynamics of the pinch\cite{giuliani2015review}, although the coupling of this motion to the circuit is outside the scope of this paper.

\subsection{Understanding the hard conductor through flux conservation}

Although the total current running through the conductor is not conserved, we can see from Eq.~(\ref{eq:Ialpha1}) that the enclosed current depends only on the initial current distribution, and the mapping between the initial and final position of the fluid element $\tR_0 \rightarrow \tr$, but not on the particular trajectory itself.
This turns out to be true for general velocity profiles, and can be traced back to the conservation of the magnetic flux.
The conservation of the flux can also help us to understand why the current behaves the way it does.

\begin{figure}[b]
	\includegraphics[width=\linewidth]{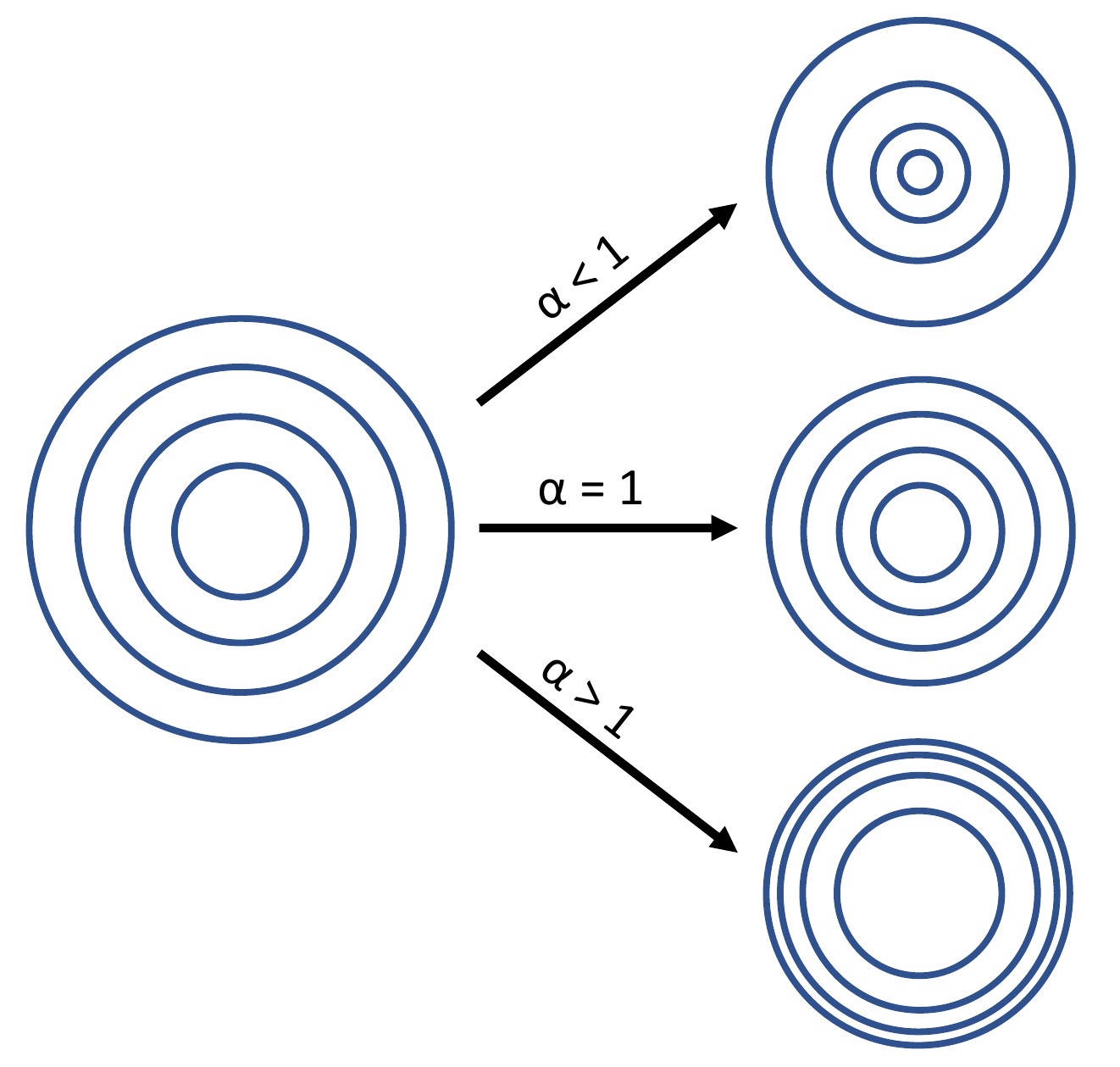}
	\caption{Schematic showing evolution of magnetic field lines in $r-\theta$ plane for sub-metric ($\alpha < 1$), metric ($\alpha = 1$), and super-metric ($\alpha > 1$) compression of the conductor.
		When $\alpha < 1$, the field gets more concentrated on axis, where it is cheaper (in terms of current) to produce, and thus the total current decreases. 
		The reverse occurs for $\alpha > 1$.
		Metric compression represents a very special case where the total current is conserved.}
	\label{fig:fieldLinesAlpha}
\end{figure}

In a cylindrically symmetric system, the normalized magnetic flux is given by
\beq
\Phi(\tr) = \int_0^{\tr} B(\tr') d\tr'. \label{eq:fluxIntegral}
\eeq
In metric compression, the magnetic field everywhere increases by the same factor as the radial coordinate decreases, so that the integral maintains the same structure.
However, this is not true when $\alpha \neq 1$.
In particular, for super-metric compression, the current density and magnetic field move closer to the conductor edge (Fig.~\ref{fig:fieldLinesAlpha}).
Because $I \propto rB$, the magnetic field is more ``expensive'' (in terms of current) to produce at larger radius, and so the total current in the conductor goes up to keep the total flux constant.
Meanwhile, for sub-metric compression, the current distribution moves in, where the magnetic field is ``cheaper'' to produce, and thus the total current goes down.

\subsection{Importance of the boundary condition}

This global non-conservation of the current for the bounded conductor arises from the fact that $v_r$ does not go to 0 at the conductor edge.
If $v_r$ went to 0 (or $\alpha$ to 1, with $j_z=0$) in an extended region at the boundary, then from Eq.~(\ref{eq:dIdt}), $dI/dt = \partial I / \partial t = 0$ at that point.
Thus, if there is a stagnant region of ideally conducting plasma outside of some radius, the current enclosed by that region will remain constant.

In a gas puff Z pinch the occurrence of deflagration processes is very likely. The initial Paschen breakdown will tend to occur at some radius $r_p$, as was found by Giuliani et al. \cite{giuliani2014temperature} in their attempt to fit the stagnation data from another experiment at the Weizmann \cite{kroupp2011ion, osin2011instabilities}.
Then, an ionization wave moves outward, meaning a plasma is continually initialized at rest at the pinch boundary \cite{golingo2005deflagration}.
Thus, a boundary condition of $v = 0$ at the conductor boundary seems somewhat reasonable.
It is therefore worthwhile to examine  velocity profiles which go to 0 at the pinch boundary.
This will in general require numerical solutions.

We note in passing that this deflagration wave also provides a non-resistive mechanism for current channel broadening, which helps to provide the broad initial current profiles.
Deflagration creates new plasma at the outer boundary of the conducting plasma region, and the current generator can only add current (in excess of plasma self-induction) at the outer edge of the plasma.
Thus, excess current is continually added to the expanding outer edge of the plasma, producing a broad initial current profile even in the absence of magnetic diffusion.

\subsection{Current evolution in the lab frame for constant velocity profiles}

So far, we have been focused on the evolution of the enclosed current in the fluid element frame.
The analysis in this frame is relatively simple, since the gain or loss of current do not depend on the specific current profile in question.
In addition, this frame naturally describes the gain or loss of current ``from the plasma.''

In an experimental setting, however, one measures the magnetic field as a function of space, not as a function of fluid element. 
Thus we must relate the current evolution in the fluid frame to the current evolution in the lab frame.

While in general there is not much progress to be made once the convective derivative is included, we can find some valuable intuition in the special case where the velocity profile is constant in time.
Then, we can use
\beq
	\frac{dv_r}{dt} = \lp \cancel{\pa{}{t}} + v_r \pa{}{r} \rp v_r
\eeq
to eliminate the $r$ derivative in Eq.~(\ref{eq:dIdt}), yielding:
\begin{align}
	\frac{dI}{dt} &= -I \pa{v_r}{r} + \frac{I}{r} v_r\\
	\frac{1}{I} \frac{dI}{dt} &= -\frac{1}{v_r} \frac{dv_r}{dt} + \frac{1}{R} \frac{dR}{dt}. \label{eq:amnonIntermediate}
\end{align}
In the second line, we have used $v_r = dR/dt$ to pass the last term from the lab to the fluid frame.
Then, we can recast Eq.~(\ref{eq:amnonIntermediate}) as a conservative equation in the fluid frame:
\begin{align}
	\frac{d}{dt} \lp \frac{I v_r}{R} \rp = 0.
\end{align}

The conservative equation allows us to easily write the lab-frame solution $I(r,t)$:
\begin{align}
	\frac{I v_r}{r} = \frac{I_0(R_0) v_r(R_0)}{R_0},
\end{align}
Here, $I_0(R_0)$ is the initial enclosed current evaluated at the starting position $R_0(r,t)$ of the fluid element that is at $r$ at time $t$.

We can examine the time-evolution of the enclosed current by taking a partial time derivative:
\begin{align}
	\frac{v_r}{r}\pa{I}{t} = \pa{}{R_0} \lp \frac{I_0(R_0) v_r(R_0)}{R_0} \rp \pa{R_0}{t}.
\end{align}
As before, we can gain insight by examining power law profiles.
Taking $v \propto r^{\alpha}$ and $I_0 \propto R_0^\beta$, we have
\begin{align}
	\sgn \lp \pa{|I|}{t} \rp = -\sgn\lp \alpha + \beta - 1\rp \sgn\lp v_r \rp,
\end{align}
where we have noted that $\partial R_0 / \partial t$ has the opposite sign of $v_r$, since it is a backwards-integration of the velocity trajectory.

Consider again the case of a compressing plasma.
If the current density is 0 in some region outside a region of current then in this region $\beta = 0$, and there will be a local decrease of enclosed current for $\alpha < 1$, as in the fluid frame.
However, if we have a finite current density, $\beta > 0$, and thus there will be a \emph{lab-frame} decrease of current only if $\alpha < 1-\beta$.
The convective derivative, which carries current density along with the plasma, tends to increase the enclosed current and make it harder for the local current to decrease.
In the important case of uniform current density, $\beta = 2$, and there is only lab-frame current decrease if $\alpha < -1$.

\begin{figure*}
	\includegraphics[width=\linewidth]{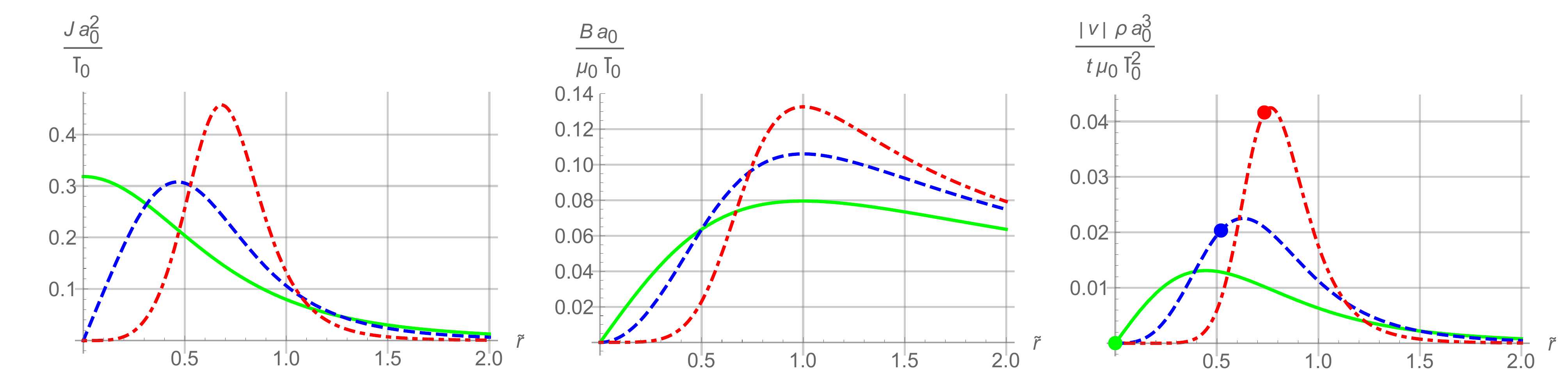}
	\caption{The function family $j_{m}(\tr)$ and $B_{m}(\tr)$, and associated short-time velocity profile $v_m(\tr)$ for different values of $m$. Different lines correspond to $m=0$ (green, solid), $m = 1$ (blue, dashed), and $m=4$ (red, dash-dotted). The maximum $B$ field always occurs at $\tr=1$. Increasing $m$ corresponds to increasing peakedness of the current profile.
	The dot on each profile indicates the point where $\partial (\tv / \tr) / \partial \tr = 0$, i.e. the transition point between super-metric and sub-metric compression.}
	\label{fig:jBvprof}
\end{figure*}

\section{Numerical solutions from flux conservation for more general scenarios} \label{sec:numerical}

Because the magnetic flux is conservered, for a comoving fluid element with radial coordinate $R(t)$,
\beq
	\frac{d}{dt} \Phi (R(t)) = 0.
\eeq
This conservation property gives us a quick way to find numerical solutions to the induction equation for more complicated, or even time-dependent, velocity profiles.
Given a velocity profile $\tv(\tr,\tdt)$, we start by numerically solving for the motion of the fluid element $\tR(\tdt)$.
We will then have a set $(\tR(\tdt),\Phi(\tR(\tdt))$, which can be interpolated to yield $\Phi(\tr,\tdt)$ over the domain of dependency of our initial fluid elements.
$\Phi$ can then be differentiated to yield $B$ and $I$.

Access to numerical solutions allows us to consider plasmas with time-dependent behavior, such as an inflow followed by an outflow, as well as plasmas with different behaviors in different regions.
Such features are essential for modeling the behavior of the annular current distributions embedded within larger conducting regions, which could characterize gas-puff Z pinches.

\begin{figure*}
	\includegraphics[width=\linewidth]{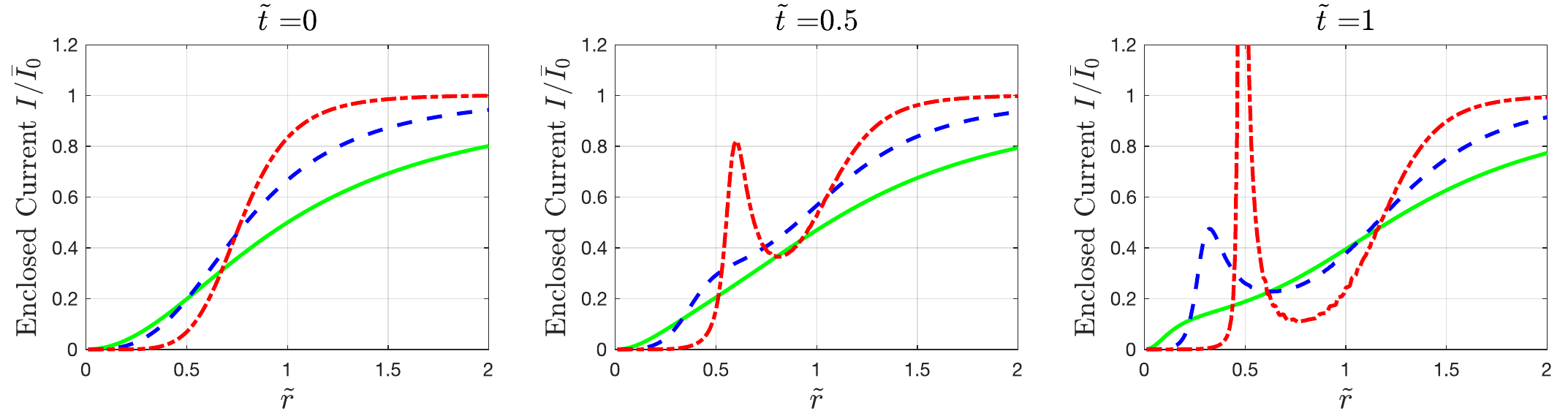}
	\caption{Evolution of the enclosed current $I(\tr) / I_0$ in the lab frame as a function of time $\tdt$ for the current annulus scenario, with a velocity profile given by Eq.~(\ref{eq:vAnnulus}).
		Line colors and styles correspond to the matching profiles in Fig.~\ref{fig:jBvprof}.
		Because the enclosed current goes from increasing at small radii to decreasing at large radii, an inverse current channel forms.}
	\label{fig:annulusEvolution}
\end{figure*}

\subsection{Annular current implosions} \label{sec:annulus}

If we have a total current $\bar{I}_0$ flowing through a uniform plasma of characteristic radius $a_0$, we can approximate the initial conditions via the one-parameter function family:
\beq
j_{m} (\tr) \equiv \frac{m + 2}{2\pi} \frac{\bar{I}_0}{a_0^2} \lp \frac{ (m + 1) \tr^m}{\lp 1 + (m+1)\tr^{m + 2}\rp^2} \rp  .
\eeq
The associated magnetic field is given by $B_\theta = \frac{1}{r} \int_0^r r j(r) dr$, which gives:
\beq
B_{m} (\tr) \equiv \frac{ \mu_0 \bar{I}_0}{2\pi a_0} \lp \frac{ (m + 1) \tr^{m+1}}{1 + (m+1)\tr^{m + 2}} \rp .
\eeq
In this family of functions, shown in Fig.~\ref{fig:jBvprof}, the maximal magnetic field is always at $\tr = 1$, and the parameter $m$ determines the peakedness of the current distribution, with higher peaking at high $m$.
Note that $j(r)$ diverges at $r = 0$ for $m < 0$.

If the plasma is initially uniform, at small times, the velocity profile will be proportional to the force density profile of the $\jvec \times \Bvec$ force, i.e.
\beq
	v_m = -\frac{m + 2}{4 \pi^2} \frac{\mu_0 \bar{I}_0^2}{ \rho a_0^3} \lp \frac{ (m + 1)^2 \tr^{2m+1}}{\lp 1 + (m+1)\tr^{m + 2}\rp^3} \rp t .
\eeq
This velocity profile is shown in Fig.~\ref{fig:jBvprof}.
Although it will only be valid at short times, the general shape is likely to persist for a longer time.
In nondimensional form, with $\tdt = t/\tau$:
\begin{align}
\tilde{v}_m &= -\lp \frac{ (m+2)(m + 1)^2 \tr^{2m+1}}{\lp 1 + (m+1)\tr^{m + 2}\rp^3} \rp \tdt \label{eq:vAnnulus}\\
\tau &= \sqrt{\frac{4\pi^2 \rho a_0^4}{\mu_0 \bar{I}_0^2}}.
\end{align}

We can apply our intuition from Section~\ref{sec:analytic} to the velocity profile in Eq.~(\ref{eq:vAnnulus}).
At low radii, the profile is clearly super-metric for $m>0$, so that the current enclosed by a fluid element is increasing, in accordance with Eq.~(\ref{eq:dIdt}) and Table~\ref{tab:dIdt}.
However, at large radii, the compression becomes sub-metric, and so the current enclosed by a fluid element is decreasing.
Thus, the current density must be decreasing between these regions, and at some point, will become negative.

This reversal of the current density can be seen in the simulations in Fig.~\ref{fig:annulusEvolution}, which show the enclosed current profile resulting from the initial conditions in Fig.~\ref{fig:jBvprof}.
For higher $m$, i.e. a more peaked initial current distribution, it occurs both earlier (due to the larger velocities) and further out.
We also see that the enclosed current at certain radii can actually exceed the enclosed current at the boundary, implying the possibility of creating a current channel stronger than that suggested by a boundary measurement, which is screened from conventional (edge) diagnostics by a reverse current.

Of course, these solutions do not represent self-consistent dynamics; as the current distribution changes, the plasma motion will change in response.
In particular, the formation of reverse currents will result in outward forces on the plasma, which will dramatically impact the dynamics.
Nevertheless, the solutions indicate an interesting possible consequence of the fact that the plasma is not a fixed, uniform conductor with a well-defined boundary, and are worthy of further investigation.

\begin{figure*}
	\center
	\includegraphics[width=\linewidth]{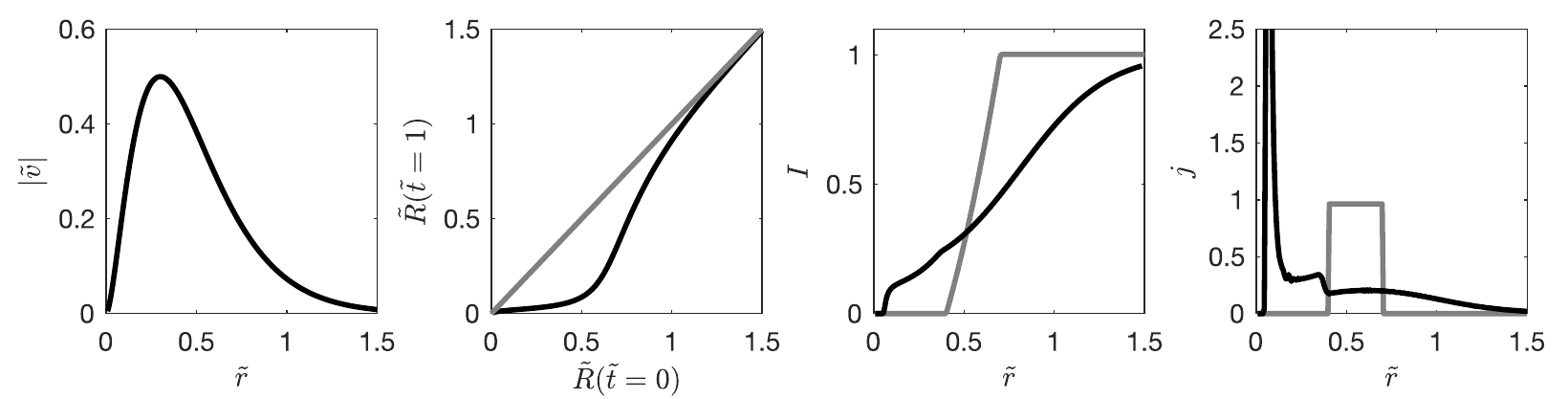}
	\caption{Details of a pre-stagnation current escape scenario.
		From left to right: velocity profile, final fluid element position as a function of initial fluid element position, enclosed current profile, and current density profile.
		All plots except the second represent profiles in the lab frame.
		The initial enclosed current distribution and current density are shown in gray.
		The current channel broadens, with the majority of the current channel moving outwards.}
	\label{fig:stagnation}
\end{figure*}

\subsection{Induction near stagnation} \label{sec:stagnation}

As the implosion progresses, Ohmic heating and compression of the plasma will rapidly increase its thermal pressure. The pressure force will often grow much larger than the magnetic force and may further accelerate the leading edge of the plasma annulus while decelerating the rear edge.
In this case, it is reasonable to assume that the current channel is trailing the region of peak compression velocity.
Thus, as stagnation approaches, the current channel will find itself in a region of sub-metric compression, where we can expect the enclosed current to decrease.

To see this in action, consider a velocity profile of the form
\beq
	\tv = d \, \tr^a e^{-\tr^b/c}.
\eeq
In terms of the coefficients $(a,b,c,d)$, the radius of maximum velocity is given by $\tr^* = (ac/b)^{1/b}$, and the maximum velocity is given by $\tv^* = d(ac/eb)^{a/b}$.
Thus, we can choose a desired general shape by choosing $a$ and $b$, and then solve for $c$ and $d$ in terms of our target $\tr^*$ and $\tv^*$.
Choosing $\tv^* = 1/2$, $\tr^*=0.3$, $a=1.7$, and $b=1$ results in the velocity profile shown in the leftmost plot Fig.~\ref{fig:stagnation}.
The corresponding change in the position of the fluid elements is shown in the second plot from the left.

We examine the effect of this velocity profile on a current annulus initially located outside of the peak inflow, from $0.4 < \tr < 0.7$.
Overall, the current channel widens dramatically as a result of the plasma motion.
However, the bulk of the current channel--roughly the outer 70\%--moves outwards.
Because the current channel is often defined as an inter-quantile range of the total current--e.g. the region between 20\% and 90\% of the peak current--a laboratory observer would describe this as outward motion of the current channel, which occurs as a result of the inward motion of the plasma.

While it is clear that the majority of the current channel can move outward under plasma compression, the inner edge of the current channel always has to move inward.
To see this, we can consider two cases: the case where there is finite current density at the origin, and the case where there is not.
If there is finite current density at the origin, then we simply observe that compression must be metric or super-metric at $r\rightarrow 0$; otherwise, there is infinite divergence in the velocity field.
Thus, the current must be increasing in time (or constant, if there is an empty region in the current profile), but cannot decrease.
If there is no current density at the origin, then there must be a location where $v_r$ is nonzero, but $I$ is 0. 
At this point, the convective term $v_r \partial I/\partial r$ will be larger in magnitude than the RHS of Eq.~(\ref{eq:dIdt}), since $I$ must go to 0 faster than its derivative.
Then, Eq.~(\ref{eq:dIdt}) says that $\partial I / \partial t = -v_r \partial I / \partial r$, so that for inward-flowing plasma, the enclosed current will grow.
Thus, during compression, the main channel can only move outward through broadening of the total channel, with inward motion of the channel at the inner edge.

The velocity profile in Fig.~\ref{fig:stagnation} can also give us intution into the behavior post-stagnation.
When a Z pinch plasma stagnates, the plasma begins to expand from the center outward.
Thus, the velocity profile will correspond to sub-metric expansion, i.e. we will have Fig.~\ref{fig:stagnation}, but with opposite sign.
Thus, by exchanging the roles of the initial and final profiles in the figure, we can gain insight into the post-stagnation behavior.

Post-stagnation, then, we see that the current channel narrows, and the majority of the current channel can be made to move inward, depending on the velocity profile. 
Just as inflow can lead to an outward motion of the main current channel, outflow can lead to inward motion.

These examples demonstrate some of the unintuitive behaviors contained in the induction equation, for even fairly straightforward velocity profiles.

\section{Comparison to Experiment} \label{sec:experiment}

Observing these unintuitive effects requires the measurement of the radial magnetic field profile.
Detailed measurements of the evolution of the radial magnetic field profile have been recently obtained at a small-scale Z pinch at the Weizmann Institute of Science. Being part of a PhD thesis\cite{stollberg2019phd}, the detailed setup and results are currently in preparation for publishing elsewhere. Here, polarization spectroscopy (simultaneous detection of the $\sigma$+ and $\sigma$- Zeeman components \cite{golingo2010note, doron2014bfield}) along with utilization of a pronounced charge state separation, as has been done previously at the Weizmann Institute \cite{rosenzweig2017measurements}, yielded the highly temporally and radially resolved magnetic field profiles.

The Weizmann plasma has a density that ranges from $10^{17}-10^{19}$ cm$^{-3}$ and a temperature that ranges from $5-20$ eV.
Thus, the electron-ion collision frequency $\tau_{ei} < 20$ ps, meaning that the electron and ion velocities extremely well equilibrated on the $30$ ns dynamical timescales considered.
A minimum value for the ion-neutral collision frequency occurs in the low-temperature limit at the low density, where the collision time in seconds is given by\cite{lieberman2005principles}:
\beq
	\tau_{ni} = 1.1 \times 10^{9} \frac{1}{n_i} \lp \frac{\alpha_R}{A_R} \rp^{1/2} Z_i, 
\eeq
where $n_i$ is the ion density in cm$^{-3}$, $\alpha_R$ is the relative polarizability of the atom, equal to $5.4$ for oxygen (used in the current experiment), and $A_R$ is the reduced mass in a.m.u., equal to 16.
Thus, the minimum neutral-ion equilibration time in low-density, singly-ionized regions of the plasma is $\tau_{ni} = 6.4$~ns.
In practice, the equilibration time will tend to be even shorter than this, since the less dense areas of the plasma tend to be hotter, and thus more highly ionized.

If the plasma were purely ideal, we could infer an average velocity profile between measurement timepoints directly from the magnetic field profile.
To do this, we take advantage of the conservation of flux.
Because the flux is conserved, if we measure the flux as a function of radius at two timepoints $t_1$ and $t_2$, we have:
\beq
\Phi(r,t_1)|_{r = R(t_1)} = \Phi(r,t_2)|_{r = R(t_2)}.
\eeq
Thus, we have a natural mapping $R(t_1) \rightarrow R(t_2)$, which corresponds to the average fluid motion over the interval $[t_1,t_2]$.

A complication is that the plasma in this experiment is also resistive.
The relative importance of advective vs resistive effects in the induction equation is given by the magnetic Reynolds number,
\beq
	\text{Rm} = \frac{\mu_0 L v}{\eta}.
\eeq
In the Weizmann experiment, the length scale is several millimeters, the implosion times are on the order of 100 ns, and the resistivities are on the order of 30 $\Omega \mu$m.
Thus, Rm $\sim 0.5$, though this number can vary significantly across the plasma.

The fact that the magnetic Reynolds number is order 1 means that resistivity and advection will be equally important in determining the magnetic field evolution.
Being in this intermediate regime will make it difficult to quantitatively model the plasma without a detailed numerical simulation incorporating both effects, which is outside the scope of this paper.
However, we can still look for qualitative signatures of the inductive effects.

\begin{figure}
	\center
	\includegraphics[width=\linewidth]{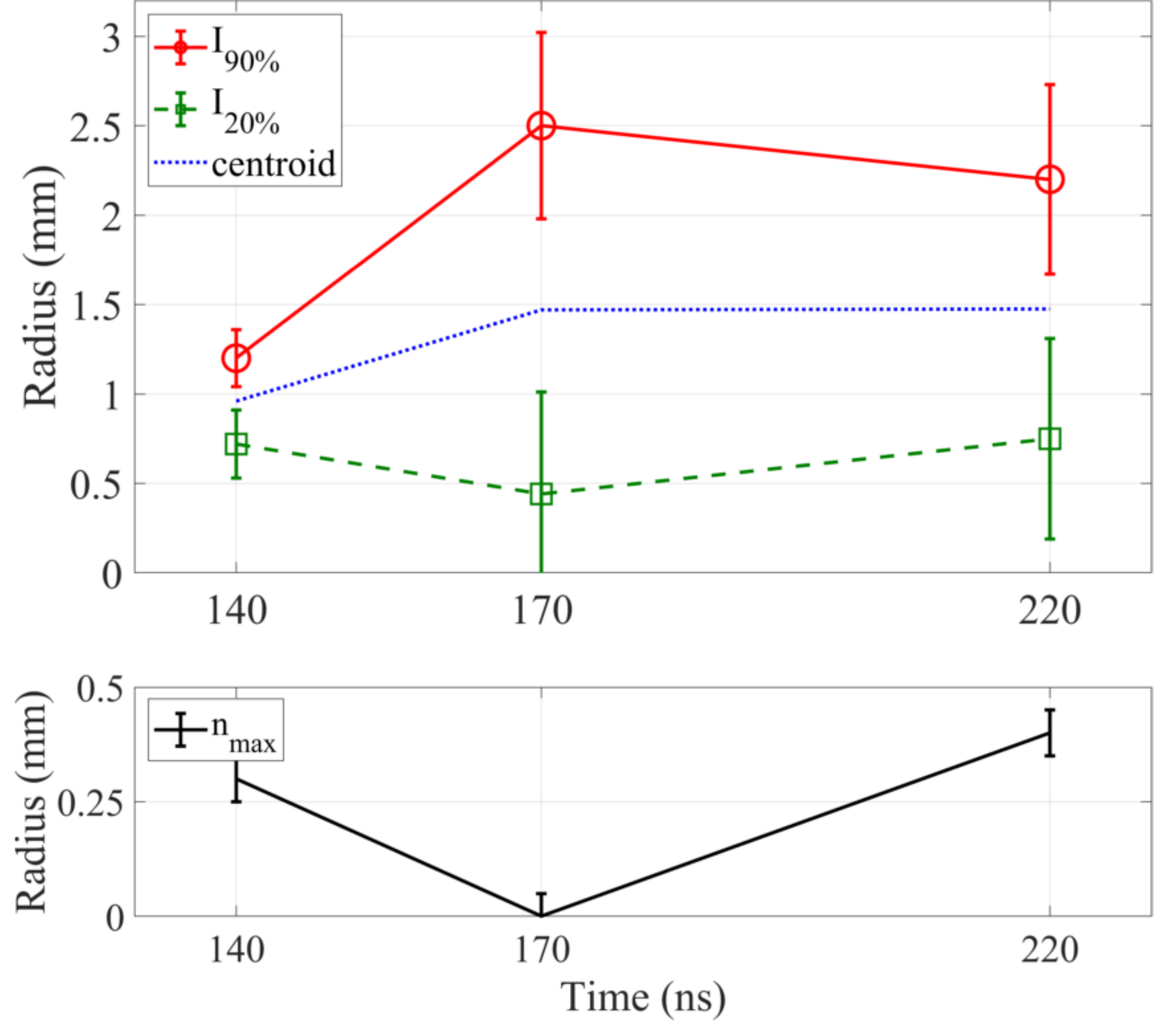}
	\caption{Upper graph: Evolution of the current channel, defined as the region between $20\%$ and $90\%$ of total current, on the Weizmann experiment around stagnation. Error bars have been evaluated individually for each point according to the signal to noise ratio of the measurement and the present current profile.
	Lower graph: Trajectory of the peak plasma density at the same times. 
	The error bars represent the spatial resolution of the setup.
	Together, the graphs show that the current channel broadens while the plasma compresses and re-contracts when the plasma expands.
	}
	\label{fig:currentEscapeStollberg}
\end{figure}

A typical trace of the current profile evolution in the Weizmann experiment around stagnation is shown in Fig.~\ref{fig:currentEscapeStollberg}.
The current channel is defined as the region between $20\%$ and $90\%$ of the total current and its centroid is at  $r_C = (r(I_{20\%})  + r(I_{90\%})) / 2$.
At 140 ns (after the start of the discharge current) the current channel, i.e. its centroid,  reaches the smallest radius and commences stagnation. 
Between 140 ns and 170 ns, the plasma continues to move inward, as indicated by the trajectory of the peak electron density in the lower plot of Fig.~\ref{fig:currentEscapeStollberg}. While the inner edge of the current channel, $I_{20\%}$, implodes together with the plasma, a distinct spread of the outer edge of the current channel, $I_{90\%}$, to much higher radii is observed. The broadening of the current channel is attended by an outward motion of its centroid. Subsequently, a narrowing of the current channel takes place, while the plasma itself expands.

Although this expansion and subsequent re-contraction of the current profile seems unintuitive on the first view, we were able to explain the behavior of the near-stagnation plasma seen in Ref. \cite{stollberg2019phd} based on a plausible model for the velocity profile in the imploding plasma.
The good agreement between the experimental data in Fig.~\ref{fig:currentEscapeStollberg} and the theoretical model in Fig.~\ref{fig:stagnation} is consistent with the observation in Ref. \cite{maron2013balance}, for remarkably different experiments and energies, that the main current channel is located {\it outside} of the point of maximum velocity and that the final stagnation is driven by the implosion pressure rather than by the magnetic field.

\section{Comparison to Haines' Resistive Model} \label{sec:haines}

We showed that for certain velocity profiles, the ideal advection model predicted a decrease of the current conducted in the plasma bulk, recovering qualitatively the experimental measurements\cite{stollberg2019phd}. 
However, other models for the current profile evolution exist as well.
For instance, one of the earliest and most successful models for the current distribution within a Z pinch, due to Haines\cite{haines1959inverse, culverwell1989nature, haines2011review}, also predicted a small current bounce around stagnation.
In this section, we first briefly review Haines' resistivity-based model, and then explain why it does not adequately describe the Weizmann experiment.

The model by Haines\cite{haines1959inverse, culverwell1989nature, haines2011review} mostly ignored the plasma motion (except for the possibility of metric compression), focusing instead on resistive diffusion of the magnetic field.
In this model, the approach was to solve for the current profile evolution consistent with the change in the total current trace $I_{tot}(t)$.
The current trace thus set a time-dependent boundary condition on the magnetic field at the outer edge of the conductor.

An immediate consequence of this boundary-value approach was that changes in the total current diffused inward from the outside edge of the conductor.
Haines' model thus predicted the formation of a positive current sheath at the outer boundary during the current rise time, and the formation of an inverse current sheath at the outer boundary when the discharge current decreases. 
Thus, the current channel would  first move outward as the current rose, and then move inward as it decayed.
However, there are three ways in which this model is inconsistent with the Weizmann observations.

First, Haines' model predicted that the additional current added at the boundary would propagate resistively inward, causing the enclosed current to increase at every point in the plasma.
In other words, although the current channel would move outward before stagnation, so that a smaller \emph{fraction} of the current would flow in the plasma bulk, the \emph{magnitude} of the current in the plasma bulk would increase. 
In the experiment, in contrast, the current in the plasma bulk decreased prior to stagnation.
However, this observed behavior \emph{is} consistent with the current channel evolution in the advection model, as shown in Fig.~\ref{fig:stagnation}.

Second, the resistive model failed to explain why the current escape and recontraction only occur around stagnation.
In fact, Haines' model predicted more rapid escape of the current channel at early times, since current added at the boundary would form a larger fraction of the overall current channel at times with less total current.
Instead, at the timepoints in the Weizmann experiment near stagnation (between 140 ns and 170 ns), the total discharge current only changed by around 10\%, inconsistent with the large-scale redistribution of the current channel.
In contrast, the advection model provides a natural transition from current channel steepening in the initial current-driven annulus (Section \ref{sec:annulus}), to current channel broadening and escape during the final pressure-driven compression near stagnation (Section \ref{sec:stagnation}).

Third, the resistive model predicted the formation of an inverse current sheath as the total current began to decay, whereas no inverse current sheath was observed on the Weizmann experiment after the peak current.
Admittedly, a small inverse sheath could have fallen within the error bounds of the magnetic  field measurement, with the presence of the sheath only observable through the apparent inward motion of the current channel. 

These three points taken together make the Haines model a poor candidate to explain the current escape in the Weizmann experiment.

Note that the formation of inverse currents at the plasma edge was one of the most interesting consequences of the Haines model, since their presence implied that the plasma would separate into a contracting inner plasma and an expanding outer plasma.
However, in subsequent years, inverse currents were also shown to result from the propagation of pressure shock waves from the plasma center\cite{lee2000reversed}.
Moreover, as we showed in Section~\ref{sec:annulus}, inverse currents can also arise fairly easily as a consequence of the ideal induction equation, for certain velocity profiles.
Indeed, as Eq.~(\ref{eq:dIdt}) and Table~\ref{tab:dIdt} show, whenever a conducting area outside of a current distribution is subjected to sub-metric compression or super-metric expansion, for example as a result of pressure forces, a reverse current distribution will develop.
This suggests that reverse currents might be even more ubiquitous and easy to produce than previously thought.

\section{Discussion of Plasma Observables For Future Studies} \label{sec:observables}

Because the magnetic Reynold's number Rm $\sim 1$, it is difficult to quantitatively disentangle resistive effects, such as those in Haines' model, from the ideal advective effects we considered.
In this section, we discuss the experimental observables which are most promising for differentiating the ideal versus resistive behavior of the plasma.

The unintuitive and nonconservative behavior of the current density under ideal induction implies that it is not the best observable to consider when attempting to back out the plasma dynamics.
Instead, a useful analysis technique could be to make use of the locally ideally-conserved quantities in the plasma, i.e. the magnetic flux $\Phi(r) = \int_0^r B dr'$ and the enclosed nucleus number per unit length $N(r) = \sum_{i} 2\pi \int_0^r n_i r' dr'$, where the sum is over all charge states of all atomic species.
Then, if flux is conserved, the curve $\Phi(N)$ will be constant at all timepoints.
Thus, if the magnetic field profile and density profile can be independently measured, changes in the curve $\Phi(N)$ should indicate non-1D-ideal magnetic field evolution.

However, such an approach is not without its drawbacks.
First, since the flux is an integral over radius, the error in the magnetic field measurement will be amplified at larger radii.
Second, in a small-scale, low-density gas puff Z-pinch, direct ion density measurements are infeasible. 
Thus, only the electron density can be measured directly; ion density must then be inferred from the electron density and temperature, and neutrals are generally entirely invisible.
Thus, constructing an accurate $N(r)$ is challenging, although bounds can be placed based on reasonable locations for the neutrals.

In addition to these integral conserved quantities, the quantity $B / r n_e$ is also constant along the electron trajectory in an axisymmetric ideal plasma \cite{kulsrud1988analysis}. 
In fact, the constancy of this quantity along the electron trajectory can induce fast magnetic field penetration into a dilute plasma\cite{fruchtman1991penetration}. 
The quantity $B / r n_e$ is thus another strong candidate for use with spectroscopic data, to evaluate the consistency of the plasma dynamics with ideal MHD.
However, it also requires simultaneous measurement of the density and magnetic field profiles, and thus is subject to many of the same errors as trying to measure $\Phi(N)$.

Thus far, we have only considered measurement of the magnetic field and nucleus density.
However, in the experiment performed at the Weizmann Institute, each charge state was peaked at a different radius; the spectra of these lines thus could in principle give measurements of the magnetic field (via Zeeman splitting \cite{rosenzweig2017measurements, stollberg2019phd}) \emph{and} radial velocities (via Doppler shifts \cite{foord1994particle,jones2011doppler}) at those locations. 
In the present experiment, the relatively low implosion velocities prevented the spectroscopic velocity determination. 
However, the ability to measure $v_r$ as a function of time would allow for the approximate reconstruction of the fluid element trajectories $R(R_0,t)$.
Thus, by comparing the flux curves $\Phi(R_0)$ for the moving fluid elements at different times, it should be possible to deduce the degree to which flux is conserved, \emph{without} relying on inferred ion densities.

Finally, given the many complications, carrying out numerical simulations that reproduce the  observed charge state distribution evolution, including ionization and particle motion, could provide a useful bridge between experiment and theory \cite{giuliani2014temperature}. 
Such simulations should be focused on the scenario we propose here, consistent with experimental findings\cite{maron2013balance}, in which the pinch transitions from a current-driven implosion at early times to a pressure-driven implosion (with a corresponding outward force on the current piston) at late times.

\section{Conclusion}

In this paper, we derived the consequences of the ideal induction equation for the current channel dynamics in a contracting or expanding Z pinch.
We showed that for non-metric compression, the current distribution can exhibit surprising behavior.
For a bounded-conductor model, we showed that the specific velocity profile within the imploding conductor had a dramatic effect on the evolution of the total current flowing through the conductor, suggesting that the evolution of the current distribution for a Z pinch could depend strongly on the specifics of the implosion profile.
In considering gas-puff Z pinches, we discussed how the outward propagation of an ionization wave implied the presence of a zero-velocity conductor boundary, which in turn could lead to the formation of reverse currents.
We also showed how certain velocity profiles could explain the expansion and re-contraction of the current channel around stagnation, thus explaining part of the findings observed in the Weizmann Z pinch experiment  by the spectroscopic magnetic field distribution in the imploding plasma.

\section*{Acknowledgments}
This work was motivated by the yet unpublished data of CS, obtained during her Ph.D research at the Weizmann Institute of Science, on the magnetic field evolution in a small-scale Z pinch.
The authors are very grateful to U. Shumlak for fruitful discussions.
This work was supported by NNSA 83228-10966 [Prime No. DOE (NNSA) DE-NA0003764], by NSF PHY-1506122, by the BSF 2017669, and by the Air Force Office of Scientific Research (USA)  [AFOSR No. FA9550-15-1-0391].
One author (IEO) also acknowledges the support of the DOE Computational Science Graduate Fellowship (DOE grant number DE-FG02-97ER25308).

%


\end{document}